\documentstyle[preprint,aps]{revtex}

\tightenlines

\begin{document}
\draft
\title{Statistics of level spacing of geometric resonances in random binary
composites }
\author{Y. Gu$^{1,2}$ and K. W. Yu$^2$}
\address{$^1$Mesoscopic Physics Laboratory, Department of Physics, \\
Peking University, Beijing 100871, China}
\address{$^2$Department of Physics, The Chinese University of Hong Kong, \\
Shatin, New Territories, Hong Kong, China}
\author{Z. R. Yang}
\address{Department of Physics, Beijing Normal University, \\
Beijing 100875, China}
\maketitle

\begin{abstract}
We study the statistics of level spacing of geometric resonances in the
disordered binary networks. For a definite concentration $p$ within the
interval $[0.2,0.7]$, numerical calculations indicate that the unfolded
level spacing distribution $P(t)$ and level number variance $\Sigma^2(L)$
have the general features. It is also shown that the short-range fluctuation 
$P(t)$ and long-range spectral correlation $\Sigma^2(L)$ lie between the
profiles of the Poisson ensemble and Gaussion orthogonal ensemble (GOE). At
the percolation threshold $p_c$, crossover behavior of functions $P(t)$ and $%
\Sigma^2(L)$ is obtained, giving the finite size scaling of mean level
spacing $\delta$ and mean level number $n$, which obey the scaling laws, $%
\delta=1.032 L ^{-1.952} $ and $n=0.911L^{1.970}$.
\end{abstract}

\pacs{PACS number(s): 77.84.Lf, 42.65.-k, 02.50.-r}

\vskip 5mm

\section{Introduction}

Random matrix theory (RMT) originated from dealing with the energy levels of
complex many-body quantum systems has become an independent new statistics.$%
^{1}$ Different from the standard statistics in physics, RMT focuses its
attention on the general properties of a number of stochastic ensembles with
common symmetry. The correlation and fluctuation of eigenvalues and
eigenfunctions in random Hamiltonian systems are the central issues in the
applications of RMT. Apart from the nuclear and nuclei fields, it was also
employed to study the critical statistics of disordered systems with various
complex interaction.$^{2-7}$ Recently, resonant properties of composite
materials have been studied extensively due to the large linear and
nonlinear optical responses.$^{8-12}$ In a dielectric network, there exist a
lot of geometric resonances randomly distributing in the resonant area.$%
^{13-16}$ For a specific sample, its resonance spectrum is very sensitive to
the microstructure. While, for a large number of samples given a parameter,
i.e., concentration $p$, the distribution of resonances is stable, which
implies some general features of resonance spectrum. The aim of this paper
is to study the level spacing statistics of geometric resonances.

In this work, a binary network is considered, where the impurity bonds with
admittance $\epsilon _{1}$ are employed to replace the bonds in an otherwise
homogeneous network of identical admittance $\epsilon _{2}$. The admittance
of each bond is generally complex and frequency-dependent. All the impurity
bonds construct the clusters subspace. For a binary composite, the
admittance ratio $h(=\frac{\epsilon _{1}}{\epsilon _{2}})$ of two components
has a branch cut on the negative axis when resonance happens.$^{17}$ Based
on the Green's-function formalism (GFF), the eigenvalues($s=\frac{1}{1-h}$
and $s\in \lbrack 0,1]$) of Green's-matrix $M$ are solved, the sequence of
which forms the resonance spectrum.$^{13,15}$ $M$ maps the geometric
configuration of the clusters subspace. Its solutions summarize the
geometric resonances of the network subject to the external sources and in
the quasistatic limit. The element of $M$ is defined as $M_{{\bf x},{\bf y}%
}=\sum_{{\bf z}\in C({\bf y})}(G_{{\bf x},{\bf y}}-G_{{\bf x},{\bf z}})$,
where ${\bf z}\in C({\bf y})$ means that the jointing points ${\bf z}$ and $%
{\bf y}$ belong to the impurity metallic cluster and are the nearest
neighbors. More clearly, $M_{{\bf x},{\bf y}}$ describes the interaction
between ${\bf x}$ and ${\bf y}$ and is closely related to the
``environment'' or the nearest neighbors of ${\bf y}$. In a dilute network,
there exists a large ``environment'' difference between two jointing points,
hence the elements of $M$ distribute more uncorrelatively rather than with
some symmetry. However, for a percolating network, due to the self-similar
structure, ${\bf x}$ and ${\bf y}$ have the similar ``environment'' and $M_{%
{\bf x},{\bf y}}$ is approximately equal to $M_{{\bf y},{\bf x}}$. It is
analogous to the Gaussian orthogonal ensemble (GOE), in which the elements
of the Hamiltonian matrix must satisfy $H_{m,n}=H_{n,m}$. So, for a
disordered composite, it is expected that the correlation and fluctuation of
eigenvalues of $M$ have the general features.

In the following, statistics of resonance level spacing in the disordered
binary composites is studied intensively on the unfolded scale. For one
sample, there are more than $700$ levels numerically solved by its
Green's-matrix $M$. In the unfolding procedure, we use a fit of the third
order polynomial to the data. For an arbitrary $p$, $1000$ samples, with
totally more than $700,000$ levels, are computed for the sake of the
ensemble averaging. For a definite $p$ within the interval $[0.2,0.7]$, our
numerical calculations indicate that the unfolded level spacing distribution 
$P(t)$ and level number variance $\Sigma ^{2}(L)$ have the general features.
It is also shown that the short-range fluctuation $P(t)$ and long-range
spectral correlation $\Sigma ^{2}(L)$ lie between the profiles of the
Poisson ensemble and GOE.

This paper is organized as follows. In the next two sections, for various $p$%
, level spacing distributions $P(t)$'s and level number variances $\Sigma
^{2}(L)$'s are calculated on the unfolded scale. Then, in Section IV, at $%
p_{c}$, crossover behavior of $P(t)$ and $\Sigma ^{2}(L)$ is obtained,
giving the finite size scaling of mean level spacing $\delta $ and mean
level number $n$. Finally, we summarize the main results in Section V.

\section{unfolded level spacing distribution}

In order to remove the system-specific mean level density or normalize the
resonance level spacing of different samples, unfolding procedure is
necessary. For a sample, the cumulative spectral function $C(s)$ of its
resonance spectrum is defined as$^{1}$ 
\begin{equation}
C(s)=\int_{-\infty }^{s}f(s^{\prime })ds^{\prime },
\end{equation}%
where $f(s^{\prime })=\sum_{n=1}^{N}\delta (s^{\prime }-s_{n})$ is the
spectral function of levels. When $N\rightarrow \infty $, $f(s^{\prime })$
should be smooth. $C(s)$ is the staircase function and is used to count the
number of levels. To satisfy that $f(s^{\prime })$ is smooth, $s$ is
rescaled by $\xi $ as 
\begin{equation}
C(s)=\xi +C_{fl}(s),
\end{equation}%
where $\xi $ is the smooth part of $C(s)$ and $C_{fl}(s)$ is the fluctuating
part of $C(s)$. Fig. 1 shows the small section of the measured spectrum. In
the following, we use the third order polynomial to fit the data. However,
when $p=0.1$ or $0.2$, the ninth or eleventh polynomial is not high enough
to fit the curve because of the large fluctuation of level spacing. So at
the dilute systems, numerical calculations are not accurate.

The level spacing distribution $P(t)$ is the probability density for
neighboring levels $\xi _{n}$ and $\xi _{n+1}$ having the spacing $t$. It is
used to describe the short-range spectral fluctuations. On the unfolded
scale, the function $P(t)$ and its first moment are normalized to unity, $%
\int_{0}^{\infty }P(t)dt=1$ and $\int_{0}^{\infty }tP(t)dt=1$. For the
uncorrelated or Poisson ensemble $P_{p}(t)=exp(-t)$, while for the GOE, or
Wigner-Dyson ensemble, $P_{WD}(t)=\frac{\pi }{2t}exp(-\frac{\pi ^{2}t}{4})$.
Fig. 2 displays the level spacing distributions $P(t)$'s for various $p$
within the interval $[0.1,0.7]$. In this figure, the distributions $P_{p}(t)$
and $P_{WD}(t)$ are drawn by the dashed and solid lines respectively. The
solid curves with the filled or opaque circles represent the critical
distributions $P_{T}(t)$ at the percolation threshold $p_{c}$. The inset of
Fig. 2 is used to describe the level spacing distributions $P(t)$'s when $%
p>p_{c}$. In this case, the calculations of the functions $P(t)$'s are not
very accurate because of the degeneracy of resonances. For a specific
sample, it is impossible to estimate where the next level is due to the
complex microstructure. However, for the disordered composites with a
definite $p$, when the $i$th level is measured, the spacing between the $i$%
th and $(i+1)$th levels satisfies the ensemble averaged distribution $P(t)$,
rather than the Poisson distribution $P_{p}(t)$ or GOE case $P_{WD}(t)$. It
is seen that the distributions $P(t)$'s lie between the profiles of the
Poisson ensemble and GOE for $p\in \lbrack 0.1,0.7]$. As discussed in the
introductory part, the critical $P_{T}(t)$ approaches the $P_{WD}(t)$ due to
the strong interaction among the metallic bonds. When $p$ is very small, $%
P(t)$ is close to $P_{p}(t)$ due to the weak interaction. At last, for an
arbitrary $p$, $P(t)$ lies between the functions $P_{p}(t)$ and $P_{T}(t)$.
In this figure, it is also shown the crossover behavior of $P(t)$.

In Fig. 2, we find that the probability of small spacing is much less than
the uncorrelated Poisson distribution. It means that some repulsion exists
between neighboring levels. The repulsion in Gaussion ensembles comes from
the symmetry of matrix elements. Here, the property of matrix elements is
described by the Green's-matrix $M$ of the GFF. So the repulsion of
resonance levels is caused by the interaction of metallic bonds. The
repulsion becomes stronger with decreasing $|p-p_{c}|$. For the GOE, we
notice the importance of the repulsion law $P(t)\propto {t^{\beta }}$ with $%
\beta =1$ for small spacing.$^{1}$ While for the composite materials with
various $p$, when $t\rightarrow 0.0$, $P(t)>t^{\beta }$. When $t>2.002$,
here $2.002$ is the second intersection point of $P_{p}(t)$ and $P_{WD}(t)$,
the long tails of all functions approach the values of the Poisson ensemble.

\section{Unfolded level number variance}

The above nearest neighbor level spacing distribution contains the
information of short scales about the resonance spectrum. Long-range
correlation is measured by the level number variance $\Sigma ^{2}(L)$, given
by$^{1}$

\begin{equation}
\Sigma ^{2}(L)=<C^{2}(L,\xi _{s})>-<C(L,\xi _{s})>^{2},
\end{equation}%
where $C(L,\xi _{s})$ counts the number of levels in the interval $[\xi
_{s},\xi _{s}+L]$ on the unfolded scale. The angular bracket denotes the
average over the starting points $\xi _{s}$. By the unfolding, $<C(L,\xi
_{s})>$ should be equal to $L$. Thus, in the interval of length $L$, one
expects to find the $L\pm \sqrt{\Sigma ^{2}(L)}$ levels on average. For the
Poisson spectrum without correlation, one obtains $\Sigma _{p}^{2}(L)=L$,
while for the GOE, the analytical result is 
\begin{equation}
\Sigma _{WD}^{2}(L)=\frac{2}{\pi ^{2}}(ln(2\pi L)+\gamma +1-\frac{\pi ^{2}}{8%
}),
\end{equation}%
where $\gamma =0.5772...$ is the Euler's constant. Fig. 3 shows the
numerical results of level number variances $\Sigma ^{2}(L)$'s for $p\in
\lbrack 0.1,0.7]$. The same lines and symbols are used as those in Fig. 2,
namely, the profiles of the Poisson ensemble and GOE are represented by the
dashed and solid lines, and the critical level number variance ${\Sigma _{T}}%
^{2}(L)$ is plotted by the solid line with the circles. It is obvious that
the correlation among levels is greater than the Poisson case and less than
the GOE case. We could not collect the data for $p>0.7$ due to the
degeneracy of eigenvalues. When $p=0.1$ or $p=0.2$, the level number
variance $\Sigma ^{2}(L)$ is out of the boundary of the Poisson ensemble.
The reason is that the data can not be fitted very well by the third or
higher order polynomial in the unfolding procedure. For $p\in \lbrack
0.3,0.7]$, $\Sigma ^{2}(L)$'s lie between the profiles ${\Sigma }_{p}^{2}(L)$
and ${\Sigma }_{T}^{2}(L)$. As shown in Fig. 3, the curves almost overlap
for $p=0.4$ and $p=0.6$, as well as for $p=0.3$ and $p=0.7$. So at the
percolation threshold, the crossover behavior of the level number variance
is numerically obtained.

\section{Crossover and finite size scaling at percolation threshold}

For a percolating network, level spacing distribution limited in the
interval $s\in \lbrack 0.25,0.75]$ has been investigated by Luck et. al..$%
^{14}$ Recent numerical calculations indicate that the resonances within $%
[0,0.25]$ and $[0.75,1]$ are important because the high values of the
inverse participation ratios (IPR) in those regions correspond to the
strongly enhanced optical responses.$^{16}$ In order to study the
criticality of level spacing distribution $P(t)$ and level number variance $%
\Sigma ^{2}(L)$ in a two dimensional network, the duality of level spacing
for binary model is discussed. We consider an infinite binary network with
concentration $p$. The admittance of impurity bonds and matrix bonds is set
to $\epsilon _{1}$ and $\epsilon _{2}$ respectively. We get the admittance
ratio $h=\epsilon _{1}/\epsilon _{2}$ and $s=\frac{1}{1-h}$. The resonance
spectrum is given by the set $\{{s_{1},s_{2},...s_{n}\}}$. Then the set $\{{%
t\}}$ of level spacing can be written as $\{{t_{1},t_{2},...t_{n-1}\}}$ with 
$t_{n-1}=s_{n}-s_{n-1}$. For a large network, $G$ is a typical configuration
of the concentration $p$. Binary model is invariant under the simultaneous
interchange $p\leftrightarrow {1-p}$ and $\epsilon _{1}\leftrightarrow
\epsilon _{2}$. So $G^{\prime }$ is also a typical configuration of the
concentration $1-p$ and we get the new admittance ratio $h^{\prime }=\frac{1%
}{h}$ and $s^{\prime }=1-s$. The spectrum of resonance is replaced by the
set $\{{{s_{1}}^{\prime },{s_{2}}^{\prime },...{s_{n}}^{\prime }\}}$ with ${%
s_{1}}^{\prime }=1-s_{1},{s_{2}}^{\prime }=1-s_{2}$ and ${s_{n}}^{\prime
}=1-s_{n}$. The new level spacing set $\{{t^{\prime }\}}$ is just the
original set $\{{t\}}$. So the duality of level spacing exists for the
binary model and it is the self-dual at $p=p_{c}=0.5$. The functions $P(t)$
and $\Sigma ^{2}(L)$ are computed based on the same set $\{{t\}}$ for $p$
and $1-p$.

In the calculations, we can not find the strict duality of level spacing.
One reason is that the binary sample is not large enough. Another is that
the degeneracy of resonances affects the accuracy of $P(t)$ and $\Sigma
^{2}(L)$ when $p>p_{c}$. Numerical results indicate the crossover of level
spacing distribution $P(t)$ and level number variance $\Sigma ^{2}(L)$, as
shown in Fig. 4 and Fig. 5. To characterize this behavior, for each
distribution $P(t)$, we compute the value $\eta =\frac{%
\int_{0}{}^{t_{0}}[P(t)-P_{WD}(t)]dt}{%
\int_{0}{}^{t_{0}}[P_{p}(t)-P_{WD}(t)]dt}$ with $t_{0}=0.4729$, the first
interaction point of $P_{p}(t)$ and $P_{WD}(t)$.$^{2,3}$ In this way, $\eta $
varies from $1[P(t)=P_{p}(t)]$ to $0[P(t)=P_{WD}(t)]$. We note that $\eta
_{T}=0.2162$ at the transition is close to the value $\eta _{A}=0.215$,
which corresponds to $P_{A}(t)$ at the Anderson transition.$^{18}$ As
observed in Fig. 4, $\eta (p)$ is closer to the value $1$ of the Poisson
ensemble with increasing $|p-p_{c}|$. In Fig. 3, we have found the linear
relation of $\Sigma ^{2}(L)$ with respect to $L$ as $\Sigma ^{2}(L)=\chi
\ast L$, where $\chi $ is called the spectral compressibility.$^{6}$ When
the unfolded number $L$ is larger, the value $\chi $ is $1$ for the Poisson
ensemble, while for the GOE or Wigner-Dyson ensemble, $\chi $ is
approximately equal to $0$. For various $p$, the slopes $\chi $'s are
plotted in Fig. 5. At $p_{c}$, $\chi =0.395$ is close to 0, rather than the
value $1$ of the Poisson ensemble. In Fig. 4 and Fig. 5, we observe the
crossover behavior of $P(t)$ and $\Sigma ^{2}(L)$, as well as the duality of
level spacing.

Finite size scaling of mean level spacing $\delta $ and mean level number $n$
are computed when the percolating sample is in size from $16\times 16$ to $%
32\times 32$. For each case, more than $700,000$ levels are calculated for
the ensemble averaging. The results are shown in Fig. 6 and Fig. 7. Finite
size scaling laws, $\delta =1.032L^{-1.952}$ and $n=0.911L^{1.970}$, are
obtained. Note that here the meaning of $L$ is different from that in Fig.
3. The scaling exponents $1.952$ and $1.970$ are universal and closely
related to the spatial dimension $(D=2)$ of the network.

\section{Conclusions}

We have investigated the fluctuation and correlation of geometric resonance
level spacing in the random binary composites by RMT. The main conclusions
include:

\begin{enumerate}
\item For a definite $p$, the unfolded level spacing $P(t)$ and level number
variance $\Sigma ^{2}(L)$ have the general features.

\item For an arbitrary $p$, the short-range spectral fluctuation $P(t)$ and
long-range spectral correlation $\Sigma ^2(L)$ lie between the profiles of
the Poisson ensemble and critical ensemble.

\item The functions $P_T(t)$ and $\Sigma _T^2(L)$ at the transition approach
the profiles of the GOE, rather than the profiles of the Poisson ensemble.

\item The crossover behavior and duality of $P(t)$ and $\Sigma ^{2}(L)$ are
found when $p$ approaches $p_{c}$.

\item At $p_c$, finite size scaling laws, $\delta =1.032L^{-1.952}$ and $%
n=0.911L^{1.970}$, are obtained.
\end{enumerate}

Statistics of eigenvalues of the Green's-matrix $M$ has been studied and the
general statistical distributions have been obtained. Statistics of
eigenvectors of $M$, which are closely related to the local electric fields,
will be published elsewhere.

\begin{figure}[tbp]
\caption{ The typical cumulative spectral function C(s) of resonance levels.
The small part of the measured spectrum is shown as the staircase function.
The smooth part $\protect\xi(s)$ is the third order polynomial whose
coefficients are found by a fit of the whole measured spectrum. }
\label{fig1}
\end{figure}

\begin{figure}[tbp]
\caption{Level spacing distribution $P(t)$ of resonance spectrum on the
unfolded scale. Here $p$ is ranged at the interval $[0.1,0.7]$. Dashed and
solid lines show $P(t)$'s of the Poisson ensemble and GOE. The solid lines
with filled or opaque symbols show the profiles of various $p$. }
\label{fig2}
\end{figure}

\begin{figure}[tbp]
\caption{Level number variance $\Sigma^2(L)$ of resonance spectrum on the
unfolded scale. Here $p$ is ranged at the interval $[0.2,0.7]$. Dashed and
solid lines show $\Sigma^2(L)$'s of the Poisson ensemble and GOE. The solid
lines with filled or opaque symbols show the profiles of various $p$.}
\label{fig3}
\end{figure}

\begin{figure}[tbp]
\caption{Dependence of $\protect\eta$ on the concentration $p$. The lines
are guides to the eyes. }
\label{fig4}
\end{figure}

\begin{figure}[tbp]
\caption{Dependence of the spectral compressibility $\protect\chi$ on the
concentration $p$. The lines are guides to the eyes. }
\label{fig5}
\end{figure}

\begin{figure}[tbp]
\caption{Finite size scaling of the mean level spacing $\protect\delta$ at $%
p_c$. The sample is in size from $16\times16$ to $32\times32$. }
\label{fig6}
\end{figure}

\begin{figure}[tbp]
\caption{ Finite size scaling of the mean level number $n$ at $p_c$. The
sample is in size from $16\times16$ to $32\times32$. }
\label{fig6b}
\end{figure}

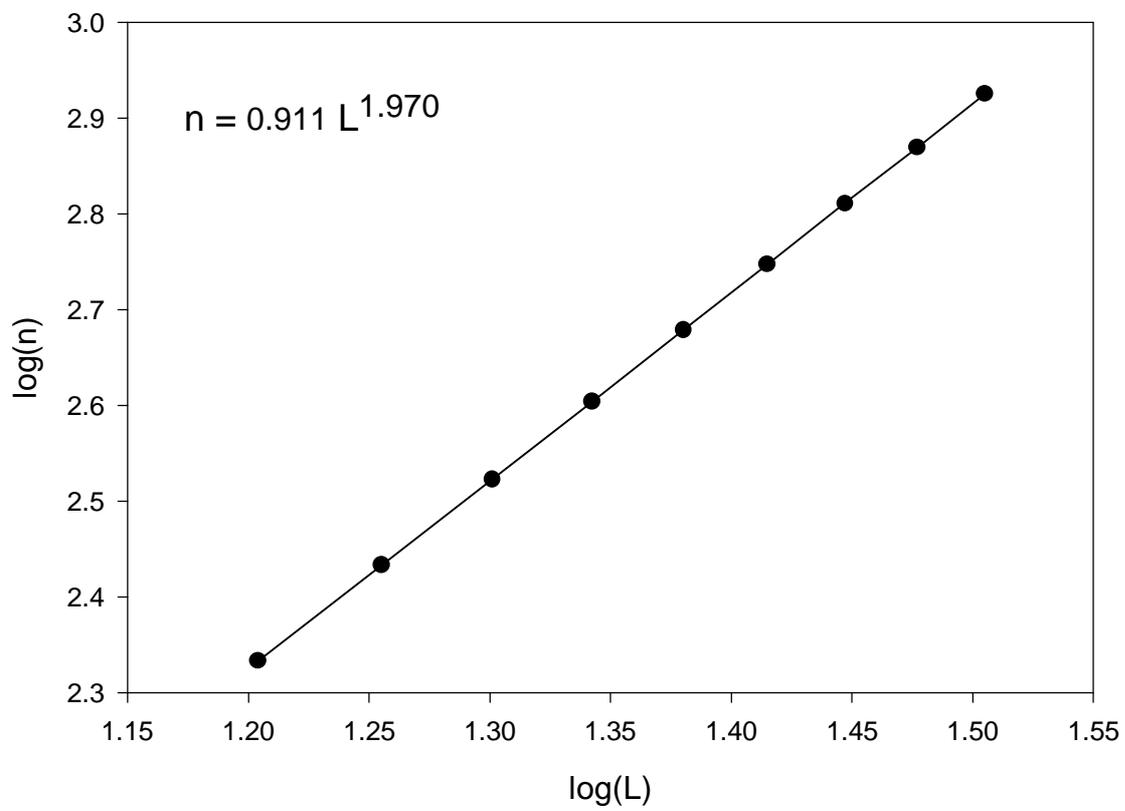
\end{document}